\documentclass[aps,pra,superscriptaddress,showpacs]{revtex4}

\usepackage{graphicx}
\usepackage{bm}

\begin{document}


\title{Nonclassical effects in cold trapped ions inside a cavity}


\author{F.L. Semi\~ao}
\email[semiao@ifi.unicamp.br]   
\author{A. Vidiella-Barranco}
\email[vidiella@ifi.unicamp.br]
\author{J.A. Roversi}
\email[roversi@ifi.unicamp.br]
\affiliation{Instituto de F\'\i sica ``Gleb Wataghin'',
Universidade Estadual de Campinas,
13083-970   Campinas  SP  Brazil}


\date{\today}

\begin{abstract}
We investigate the dynamics of a cold trapped ion coupled to the quantized field
inside a high-finesse cavity, considering exact resonance between the ionic internal 
levels and the field (carrier transition). We derive an intensity-dependent hamiltonian 
in which terms proportional to the square of the Lamb-Dicke parameter ($\eta$) are 
retained. We show that different nonclassical effects arise in the dynamics of the ionic 
population inversion, depending on the initial states of the vibrational motion/field and
on the values of $\eta$.  
\end{abstract}


\maketitle

\section{Introduction}

The manipulation of simple quantum systems such as trapped ions \cite{wine98} has 
opened new possibilities regarding not only the investigation of foundations of 
quantum mechanics, but also applications on quantum information. In such a system,
internal degrees of freedom of an atomic ion may be coupled to the electromagnetic 
field as well as to the motional degrees of freedom of the ion's center of mass. 
Under certain circumstances, in which full quantization of the three sub-systems becomes
necessary, we have an interesting combination of two bosonic systems coupled to a spin 
like system. A possibility is to place the ion inside a high finesse cavity in such a 
way that the quantized field gets coupled to the atom. A few papers 
discussing that arrangement may be found in the literature, e.g., the investigation 
of the influence of the field statistics on the ion dynamics \cite{zeng94,knight98}, 
the transfer of coherence between the motional states and the field \cite{parkins99},
a scheme for generation of matter-field Bell-type states \cite{ours01}, and even 
propositions of quantum logic gates \cite{gates}. 
Several interaction hamiltonians analogue to the ones found in quantum optical 
resonance models may be constructed even in the case where the field is not 
quantized, but having the center-of-mass vibrational motion playing the role of the field. 
For instance, if we consider a two-level atom having atomic transition 
$\omega_0$ and center-of-mass oscillation frequency of $\nu$ in interaction with a 
laser of frequency $\omega_L$, if $\omega_L-\omega_0=-\nu$ (laser tuned to the 
first red sideband), it results an interaction hamiltonian of the Jaynes-Cummings 
type
\begin{equation}
\hat{H}_i= i\hbar\Omega(\sigma_+\hat{a} - \sigma_-\hat{a}^\dagger).\label{HJCM}
\end{equation}
If $\omega_L-\omega_0=\nu$ (laser tuned to the first blue sideband), the resulting
hamiltonian is an ``anti-Jaynes-Cummings'' type, or
\begin{equation}
\hat{H}_i= i\hbar\Omega(\sigma_+\hat{a}^\dagger - \sigma_-\hat{a}).\label{HAJCM}
\end{equation}
Note that here the bosonic operators, $\hat{a}^\dagger,\hat{a}$, 
are relative to the center-of-mass oscillation motion.
The ion itself is considered to be confined in a region much smaller than the laser 
light wavelength (Lamb-Dicke regime). Other forms of interaction hamiltonians may be
constructed in a similar way \cite{vogel95,gerry97}.

In this paper we explore further the consequences of having the trapped ion in
interaction with a quantized field. We show that under certain conditions, the 
quantum nature of the field is able to induce intensity-dependent effects in
the trapped ion dynamics, somehow analogue to models in cavity quantum 
electrodynamics \cite{buck81,ours98}. We shall remark that here we derive a 
hamiltonian with an intensity-dependent coupling from a more general hamiltonian, 
which is different from the phenomenological approach discussed in 
\cite{buck81,ours98}.

\section{The model}

We consider a single trapped ion, within a Paul trap, placed inside a high finesse 
cavity, and having a cavity mode coupled to the atomic ion. The vibrational
motion is also coupled to the field as well as to the ionic internal degrees of 
freedom, in such a way that the hamiltonian will read \cite{knight98,ours01}. 
\begin{equation}
\hat{H}=\hbar\nu \hat{a}^{\dagger}\hat{a} + \hbar\omega\hat{b}^{\dagger}\hat{b}
+\hbar\frac{\omega_0}{2}\sigma_z +\hbar g(\sigma_+ + \sigma_-)(\hat{b}^{\dagger}+
\hat{b})\cos\eta(\hat{a}^{\dagger}+\hat{a}),
\label{H}
\end{equation}
where $\hat{a}^{\dagger}(\hat{a})$ denote the creation (annihilation) operators of
the center-of-mass vibrational motion of the ion (frequency $\nu$), 
$\hat{b}^{\dagger}(\hat{b})$ are the creation (annihilation) operators of photons 
in the field mode (frequency $\omega$), $\omega_0$ is the atomic frequency 
transition, $g$ is the ion-field coupling constant, and $\eta=2\pi a_0/\lambda$ is 
the Lamb-Dicke parameter, being $a_0$ the amplitude of the harmonic motion and 
$\lambda$ the wavelength of light. Note that the ion's position inside the cavity
is different from the one considered in \cite{ours01}. Tipically the ion is well
localized, confined in a region much smaller than light's wavelenght, or $\eta\ll 1$
(Lamb-Dicke regime). Usually expansions up to the first order in $\eta$ are made in
order simplify hamiltonians involving trapped ions, which results in Jaynes-Cummings like
hamiltonians such as the ones in Eqs. (\ref{HJCM}) and (\ref{HAJCM}). However, even for
small values of the Lamb-Dicke parameter, we show that in the situation here discussed, 
an expansion up to second order in $\eta$ becomes necessary. Several interesting effects, 
such as long time scale revivals will depend on those terms, as we are going to show. 
We may expand the cosine in Eq. (\ref{H}) and obtain
\begin{equation}
\cos\eta(\hat{a}^{\dagger}+\hat{a})\approx 1-\frac{\eta^2(1+
2\hat{a}^{\dagger}\hat{a})}{2}-\frac{\eta^2(\hat{a}^\dagger{}^2+\hat{a}^2)}{2}.
\label{expan}
\end{equation}
The interaction hamiltonian will become
\begin{equation}
\hat{H}_i=\hbar g\left[1-\frac{\eta^2(1+2\hat{a}^{\dagger}\hat{a})}{2}
-\frac{\eta^2(\hat{a}^\dagger{}^2+\hat{a}^2)}{2}\right]
(\sigma_+ + \sigma_-)(\hat{b}^{\dagger}+\hat{b}).
\label{hamil}
\end{equation}
We may then rewrite the hamiltonian above in the interaction picture, in order to 
apply a rotating wave approximation. If we tune the light field so that it exactly
matches the atomic transition, i.e., $\omega_0-\omega=0$ (carrier transition), and 
after discarding the rapidly oscillating terms, we obtain the following 
(interaction picture) interaction hamiltonian:
\begin{equation}
\hat{H}^I_i= \hbar g \left[1-\frac{\eta^2(1+2\hat{a}^{\dagger}\hat{a})}{2}\right]
(\sigma_- \hat{b}^{\dagger} + \sigma_+ \hat{b}).
\label{hamilint}
\end{equation}
The resulting hamiltonian is alike a Jaynes-Cummings hamiltonian. It describes
the annihilation of a photon and a simultaneous atomic excitation 
($\hat{b}^{\dagger}$ and $\hat{b}$ are {\it field operators}), but having an
effective coupling constant which depends on the excitation number 
($\hat{a}^{\dagger}\hat{a}$) of the ionic oscillator. As we are going to show, 
this will bring interesting consequences on the ion's center-of-mass dynamics. 
We would like to point out that we have retained terms of the order of $\eta^2$ in the
co-sine expansion, which are much smaller than one. However, the product
$\eta^2 \langle \hat{a}^{\dagger}\hat{a} \rangle$ will not be negligible for a
sufficiently large excitation number of the vibrational motion.

The evolution operator associated to the hamiltonian (\ref{hamilint}) is given by 
\begin{equation}
\hat{U}(t)=\hat{C}_{m;n+1} |e\rangle\langle e| + \hat{C}_{m;n} |g\rangle\langle g|
-i\hat{S}_{m;n+1}\hat{b} |e\rangle\langle g|
- i \hat{b}^\dagger \hat{S}_{m;n+1} |g\rangle\langle e|,
\label{eop}
\end{equation}
where 
\begin{equation}
\hat{C}_{m;n+1}=\cos\left(g\left[1-\eta^2(1+
2\hat{m})/2 \right]\sqrt{(\hat{n}+1)}\:t\right),\label{cnmu}
\end{equation}
\begin{equation}
\hat{C}_{m;n}=\cos\left(g\left[1-\eta^2(1+
2\hat{m})/2 \right]\sqrt{\hat{n}}\:t\right),\label{cnm}
\end{equation}
and
\begin{equation}
\hat{S}_{m;n+1}=\frac{\sin\left(g\left[1-\eta^2(1+
2\hat{m})/2 \right]\sqrt{(\hat{n}+1)}
\:t\right)}{\sqrt{(\hat{n}+1)}}, \label{snmu}
\end{equation}
where we have used the notation $\hat{m}=\hat{a}^{\dagger}\hat{a}$ and 
$\hat{n}=\hat{b}^\dagger\hat{b}$.

We consider now the following (product) initial state
\begin{equation}
\hat{\rho}(0)=|e\rangle \langle e| \otimes \hat{\rho}_f(0) \otimes
\hat{\rho}_{v}(0),
\label{initials}
\end{equation}
or ion's internal levels prepared in the excited state $|e\rangle \langle e|$,
the field prepared in a generic state $\hat{\rho}_f(0)$, and the
ion's vibrational center of mass motion prepared in a state $\hat{\rho}_{v}(0)$.
Its time evolution, governed by the evolution operator (\ref{eop}) results, at a time 
$t$, in the following (joint) state,   
\begin{eqnarray}
\hat{\rho}(t)&=&\hat{C}_{m;n+1}\hat{\rho}_f(0)\hat{\rho}_v(0) \hat{C}_{m;n+1}
|e\rangle \langle e| + \hat{b}^\dagger\hat{S}_{m;n+1}\hat{\rho}_f(0)
\hat{\rho}_v(0) \hat{S}_{m;n+1}\hat{b} |g\rangle \langle g| \nonumber \\ 
&+& i\hat{C}_{m;n+1}\hat{\rho}_f(0)\hat{\rho}_v(0) \hat{S}_{m;n+1}\hat{b}
|e\rangle \langle g| -i \hat{b}^\dagger\hat{S}_{m;n+1}\hat{\rho}_f(0)
\hat{\rho}_v(0) \hat{C}_{m;n+1} |g\rangle \langle e|, \label{densop}
\end{eqnarray}
where the operators $\hat{C}$ and $\hat{S}$ are the ones in Eq. 
(\ref{cnmu}), (\ref{cnm}) and (\ref{snmu}) above.

The state in Eq. (\ref{densop}) is, in general, an entangled state 
involving the ion's internal (electronic) degrees of freedom, the vibrational 
motion as well as the cavity field. We will now investigate the atomic population 
inversion considering simple initial states for the center-of-mass vibrational
motion and the cavity field.

\section{Atomic dynamics}

The (internal level) ionic dynamics will depend on the distributions of
initial excitations of both the field and the center center-of-mass vibrational motion, 
given by $\langle n|\hat{\rho}_f(0)|n\rangle=\rho_{n,n}^f(0)$, 
$\langle m|\hat{\rho}_v(0)|m\rangle=\rho_{m,m}^v(0)$ respectively. For
instance, the atomic population inversion may be written as
\begin{equation}
W(t)=Tr\left[ \sigma_z \hat{\rho}(t) \right]=\sum_{n,m=0}^\infty
\rho_{n,n}^f(0)\rho_{m,m}^v(0)\:\cos\left[2g\left(1-\eta^2(1+2m)/2\right)
\sqrt{n+1}\: t\right].
\label{inver}
\end{equation}

Due to the different frequencies 
($\nu=2g[1-\eta^2(1+2m)]\sqrt{n+1}$) in the co-sine argument, we expect different 
structures of beats in the inversion due to different preparations of the center-of-mass 
motion (index $m$, without a square root) and the field (index $n$, with $\sqrt{n+1}$). 
We immediately identify two well known particular cases -  if the vibrational motion is 
prepared in a number state ($\rho^v_{m,m}(0)=\delta_{k,m}$), the atomic inversion in Eq. 
(\ref{inver}) will reduce to the characteristic pattern of the Jaynes-Cummings model - 
if otherwise the field is prepared in a number state ($\rho^f_{n,n}(0)=\delta_{l,n}$), we 
have a periodic behavior for the atomic inversion (no square root in the co-sine argument). 
This is shown in Fig. 1, where the field is initially in the vacuum state and the ion 
in a coherent state of the center-of-mass motion. 
The Lamb-Dicke parameter $\eta$ is typically less than unity, and the lowest order term 
in the expansion above, [see Eq.(\ref{expan})] is of second order in $\eta$. 
Nevertheless, as we are going to show, interesting effects will arise if the 
term $\eta^2(1+2m)$ becomes large enough \footnote{In order to be able to neglect higher 
terms in the co-sine expansion, one has to keep the product $\eta^2 m$ small enough.}

We now consider both the center-of-mass motion and the field initially prepared in 
coherent states and the atom in the excited state $|e\rangle$. After performing the 
summation over $m$ in Eq.(\ref{inver}), we obtain the following expression
\begin{equation}
W(t)=\sum_{n=0}^\infty \rho_{n,n}^f(0) w_n(t),\label{invera}
\end{equation}
with
\begin{eqnarray}
w_n(t)&=&\exp\left\{-\overline{m}\left[1-\cos\left(2\eta^2\sqrt{n+1}\,gt\right)
\right]\right\}\nonumber\\
&\times&\cos\left\{2\left(1-\frac{\eta^2}{2}\right)\sqrt{n+1}\,gt-
\overline{m}\,\sin\left[2\eta^2\sqrt{n+1}\,gt\right]\right\}.\label{term}
\end{eqnarray}
The dynamics of the population inversion predicted by Eq. (\ref{inver}) may be interpreted 
in terms of two families of ``revival'' times. The revival times associated to the 
field ($t_r^f$), when terms $n$ and $n+1$ are in phase \footnote{It is convenient to
employ the ``scaled time'' $gt$ instead of $t$.}
\begin{equation}
g t_r^f \approx \frac{2 k \pi \sqrt{\overline{n}}}{1-\frac{\eta^2 
(1 + 2\overline{m})}{2}}; 
\ \ \ \ k=1,2,\ldots ,\label{revf}
\end{equation}
depend on $\overline{m}$, the mean excitation number of the center-of-mass motion, and on
$\overline{n}$, the mean excitation number of the field. On the other hand, 
the revival times associated to the vibrational motion ($t_r^v$), when terms $m$ and 
$m+1$ are in phase, depend only on $\overline{n}$, the mean excitation number of the 
field, or
\begin{equation}
g t_r^v \approx \frac{k \pi}{\eta^2 \sqrt{\overline{n}+1}}; 
\ \ \ \ k=1,2,\ldots ,\label{revv}
\end{equation} 

It is also possible to estimate collapse times in either case, or
\begin{equation}
g t_c^f \sim \frac{1}{1-\frac{\eta^2 (1 + 2\overline{m})}{2}},\label{colf}
\end{equation}
and
\begin{equation}
g t_c^v \sim \frac{1}{\eta^2 \sqrt{\overline{m}(\overline{n}+1)}}.\label{colv}
\end{equation}
We note that distinct patterns for the atomic inversion will arise depending on the 
values of $\eta^2$, $\overline{m}$ and $\overline{n}$, which are quantities determining 
the revival/collapse times. In order to appreciate those different situations, we show, 
in what follows, some plots of the atomic inversion for different excitations of the 
ionic motion and field as well as for different values of $\eta$.
 
In Fig. 2 we have a plot of the atomic inversion as a function of time, having 
$\eta=0.02$,  $\overline{m}=4$ and $\overline{n}=25$. It is shown the usual pattern of 
collapses and revivals. In this case the relevant times are so that
$g t_c^f\approx 1$ and 
$g t_r^f\approx 31$. For a larger Lamb-Dicke parameter, $\eta=0.04$, the oscillations 
in the atomic inversion are attenuated, as we see in Fig. 3. 
This means that 
one has to be careful while expanding over the parameter $\eta$, since in the model 
we are here discussing, terms of the order of $\eta^2$ have a significant influence on the 
system evolution. Another interesting behavior predicted by Eq. (\ref{inver}) is the 
phenomenon of ``super-revivals'' of the atomic inversion, or revivals taking place at long 
times. Those may occur in driven atoms \cite{sergio94} or in trapped ions under
certain conditions \cite{ours99}. Our model also predicts ``super revivals'' if both the
center-of-mass motion and the field are prepared in coherent states, as it is shown in 
Fig. 4 (same parameters as in Fig. 2). The ``super revival'' time in this case is given
by $g t_r^v\approx 1540$, with a collapse time $g t_c^v\sim 245$. Still for long times, 
if $\eta=0.04$, as it is shown in Fig. 5, there will be changes in the ``super revival'' 
time, $g t_r^v\approx 385$ as well as in the collapse time, $g t_c^v\sim 61$. 
In this case the revival time for ``ordinary revivals'' is very close to the 
collapse time associated to the long time scale dynamics, so that the revivals themselves
are attenuated (compare Fig. 2 with Fig. 3). 
Different patterns will occur for different values of $\overline{m}$ and $\overline{n}$. 
For instance, if $\overline{m}=25$, $\overline{n}=4$ 
and $\eta=0.2$, we have $g t_r^v\approx 48$ and $g t_r^f\approx 14$. In this case there is
only one oscillation in the collapse region, as shown in Fig. 6. 
A different pattern for the oscillations occurs if the collapse time $g t_c^v\sim 24$ 
is slightly less than the revival time $g t_r^f\approx 26$ (see Fig. 7). The short time
revivals are strongly suppressed, as we see in Fig. 7. We may compare the atomic inversion
in Fig. 7 to the one in Fig. 2; in the latter case the collapse time is long enough to still
allow revivals, while in the former one a shorter collapse time significantly reduces the 
amplitude of the revivals (at $g t_r^f\approx 26$). Apart from that, we also note a 
modulation in the oscillations, which is basically due to the combination of oscillations 
originated from the field and vibrational motion quanta distributions. As a final 
comment, we have that in the case of the periodic dynamics (Fig. 1), 
$g t_r^f= 0$ and $g t_r^v\approx 1257$, which is in agreement with the revival 
times shown in Fig. 1. We would like also to point out that dissipation in the cavity 
and loss of coherence in the vibrational motion will certainly have a
destructive action, and effects occurring at longer time scales may not be apparent.
However, the atomic inversion is highly sensitive to the initial state preparation 
and the Lamb-Dicke parameter also for short times, and the observation of at least some of 
the effects above described might become possible with further improvements in the 
quality of cavities as well as in the control of trapped ions \cite{blatt}.

\section{Conclusion}

We have investigated the dynamics of a single trapped ion enclosed in a cavity. We
have found that, in the case of exact atom-field resonance (carrier transition), 
terms of the order of $\eta^2$ can make a significant contribution and should therefore 
be retained. This gives rise to a hamiltonian containing an intensity-dependent coupling.
Besides, the quantum nature of the field plays an important role. 
We have shown that, the atomic inversion as a function of time may display 
different structures of beats depending on the initial preparation of the electromagnetic 
field and the ionic motion as well as on the Lamb-Dicke parameter $\eta$, and effects 
such as suppression/attenuation of the Rabi oscillations, long time scale revivals 
as well as a periodic dynamics may occur. Those interesting features may be 
understood in terms of the two collapse times and two revival times characteristic 
of the dynamics.  


%


\begin{acknowledgments}
We would like to thank Dr. M.A. Marchiolli for valuable comments.
This work is partially supported by CNPq (Conselho Nacional para o 
Desenvolvimento Cient\'\i fico e Tecnol\'ogico), and FAPESP (Funda\c c\~ao 
de Amparo \`a Pesquisa do Estado de S\~ao Paulo), Brazil, and it is linked 
to the Optics and Photonics Research Center (FAPESP).
\end{acknowledgments}


\begin{figure}
\includegraphics[width=10cm,height=7.5cm]{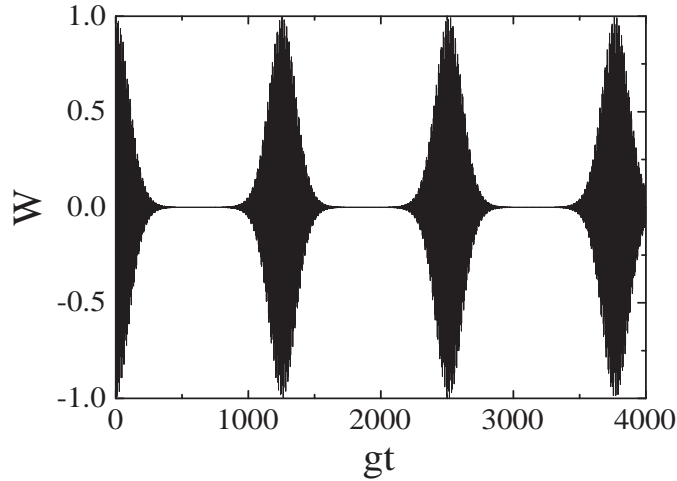}
\caption{\label{fig:fig1} Atomic inversion as a function of the scaled time $gt$. 
The field is initially in the vacuum state, $\overline{n}=0$, and the vibrational 
motion in a coherent state with $\overline{m}=4$. The ion is prepared in its internal 
excited state, and $\eta=0.05$.}
\end{figure}

\begin{figure}
\includegraphics[width=10cm,height=7.5cm]{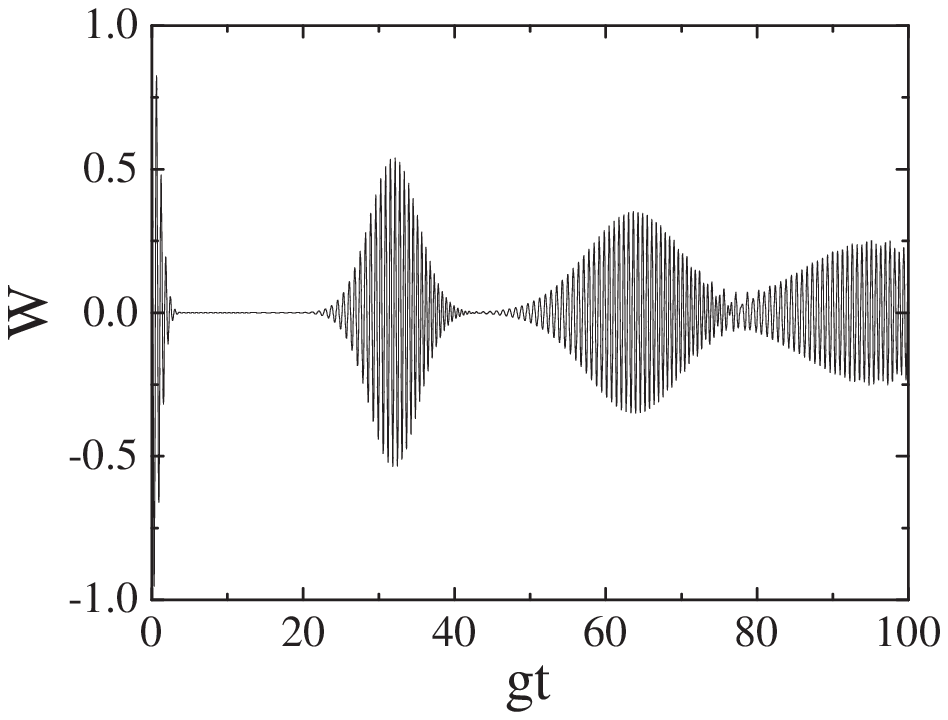}
\caption{\label{fig:fig2} Atomic inversion as a function of the scaled time $gt$. 
The field is initially in a coherent state, $\overline{n}=25$, and the vibrational 
motion in a coherent state with $\overline{m}=4$. The ion is prepared in its internal 
excited state, and $\eta=0.02$.}
\end{figure}

\begin{figure}
\includegraphics[width=10cm,height=7.5cm]{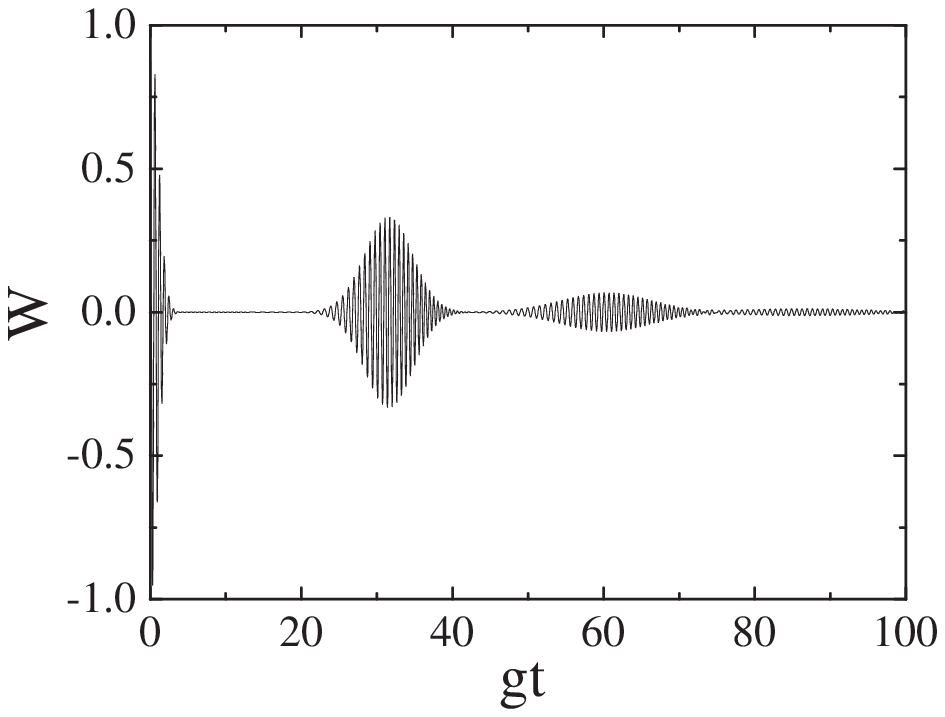}
\caption{\label{fig:fig3} Atomic inversion as a function of the scaled time $gt$. The 
field is initially in a coherent state, $\overline{n}=25$, and the vibrational motion in a
coherent state with $\overline{m}=4$. The ion is prepared in its internal excited 
state, and $\eta=0.04$.}
\end{figure}

\begin{figure}
\includegraphics[width=10cm,height=7.5cm]{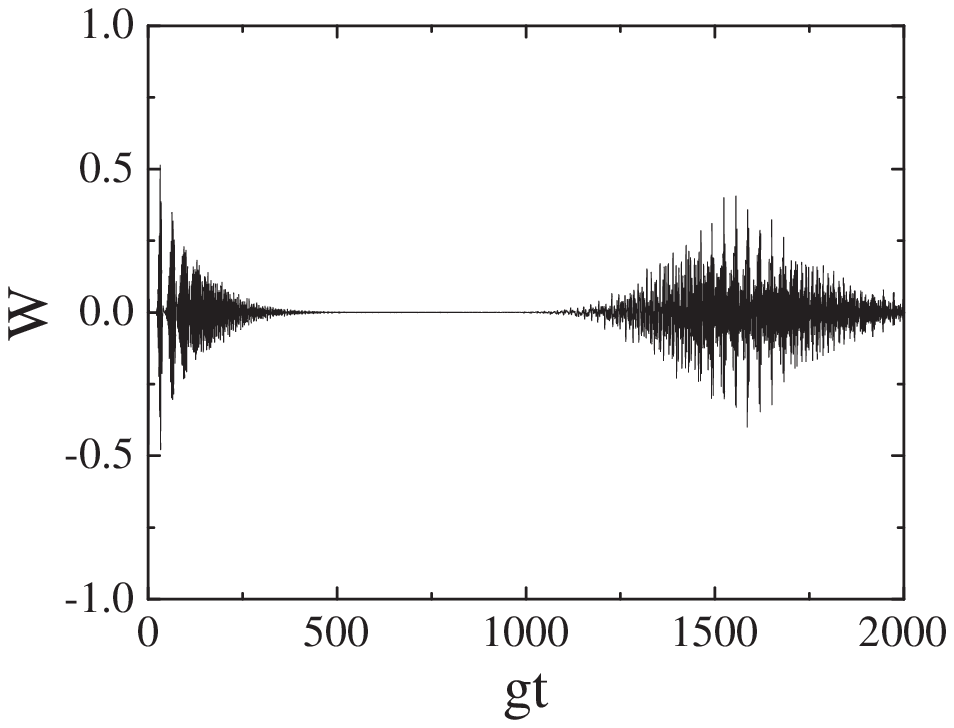}
\caption{\label{fig:fig4} Atomic inversion as a function of the scaled time $gt$. 
The field is initially in a coherent state, $\overline{n}=25$, and the vibrational 
motion in a coherent state with $\overline{m}=4$. The ion is prepared in its internal 
excited state, and $\eta=0.02$.}
\end{figure}

\begin{figure}
\includegraphics[width=10cm,height=7.5cm]{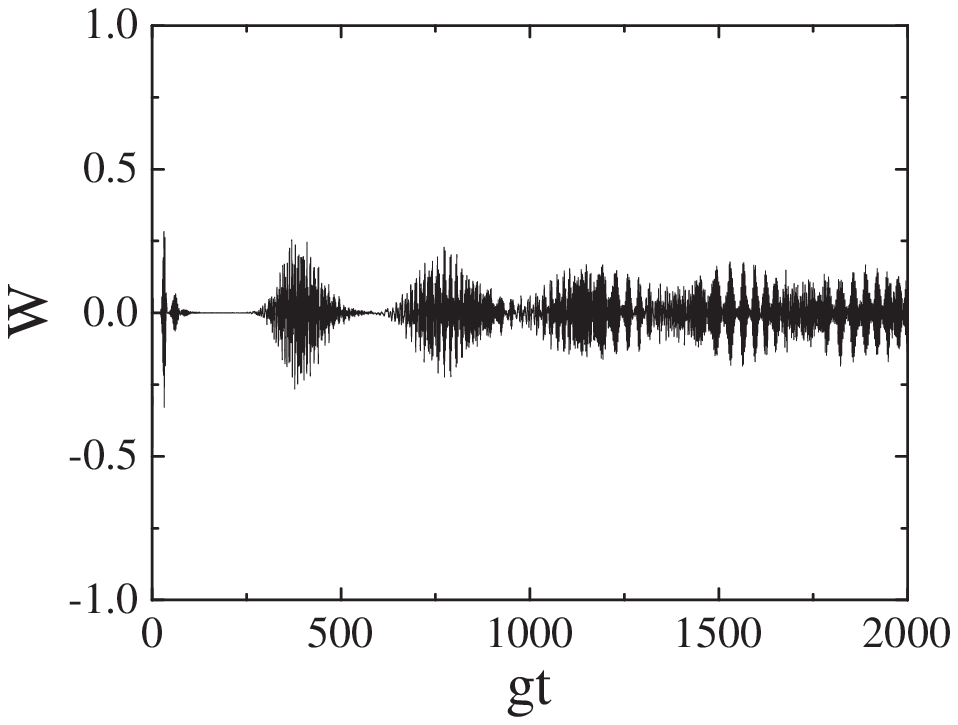}
\caption{\label{fig:fig5} Atomic inversion as a function of the scaled time $gt$. The 
field is initially in a coherent state, $\overline{n}=25$, and the vibrational motion in a
coherent state with $\overline{m}=4$. The ion is prepared in its internal excited 
state, and $\eta=0.04$.}
\end{figure}

\begin{figure}
\includegraphics[width=10cm,height=7.5cm]{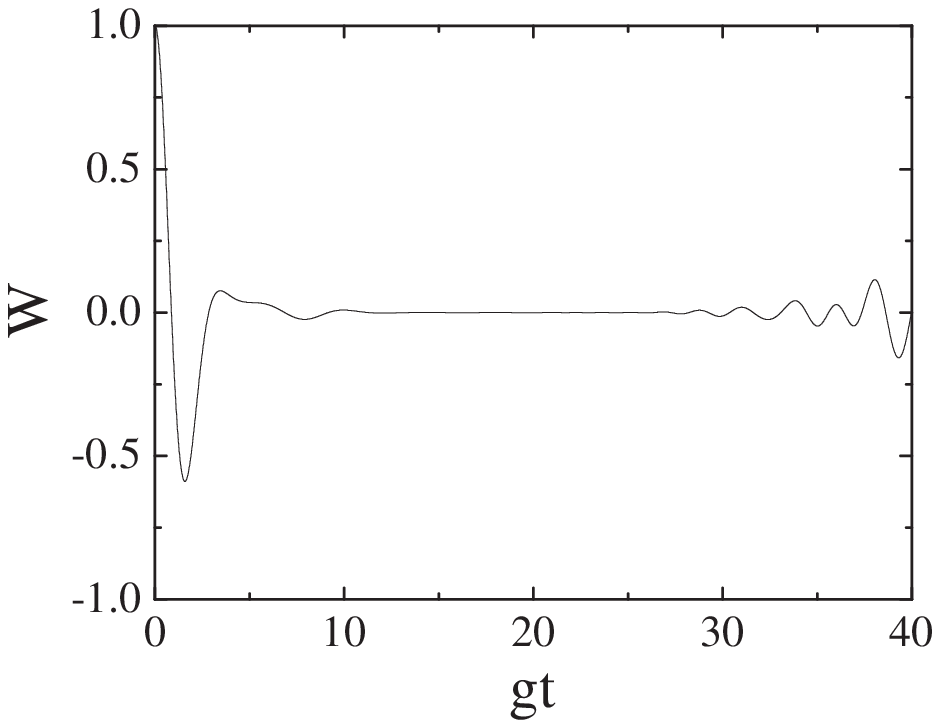}
\caption{\label{fig:fig6} Atomic inversion as a function of the scaled time $gt$. The 
field is initially in a coherent state, $\overline{n}=1.69$, and the vibrational motion in 
a coherent state with $\overline{m}=10.24$. The ion is prepared in its internal excited 
state, and $\eta=0.2$.}
\end{figure}

\begin{figure}
\includegraphics[width=10cm,height=7.5cm]{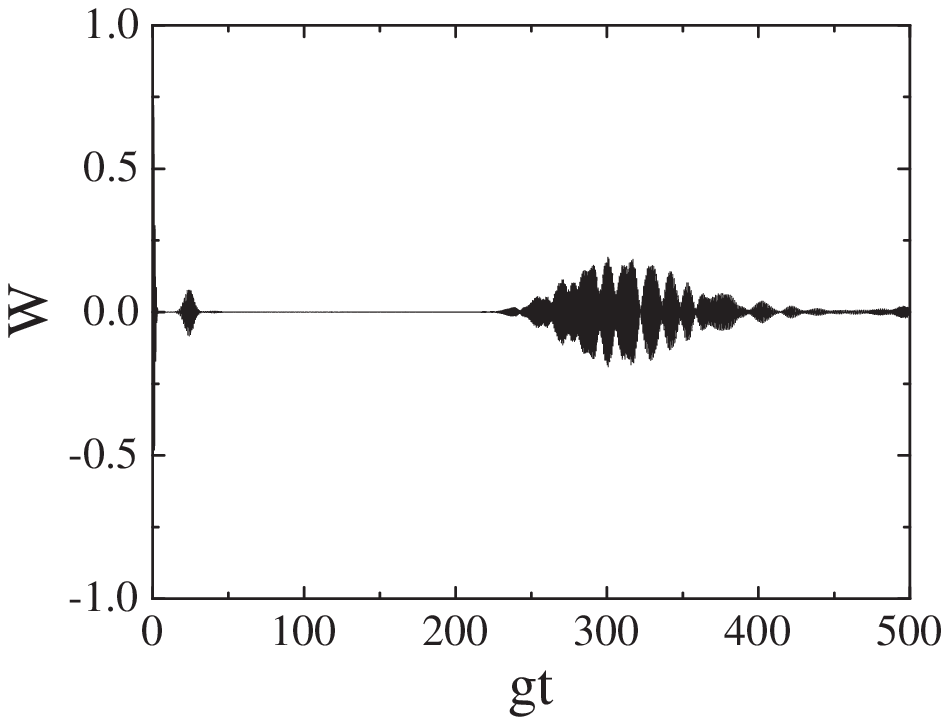}
\caption{\label{fig:fig7} Atomic inversion as a function of the scaled time $gt$. The 
field is initially in a coherent state, $\overline{n}=16$, and the vibrational motion 
in a coherent state with $\overline{m}=16$. The ion is prepared in its internal excited 
state, and $\eta=0.05$.}
\end{figure}

\end{document}